\documentclass[11pt,a4paper]{article}
\usepackage{amsthm,amssymb,epsfig,latexsym}
\newcommand{\be}[1]{\begin{equation}\label{#1}}
\newcommand{\ba}[1]{\begin{eqnarray}\label{#1}}
\newcommand{\ee}{\end{equation}}
\newcommand{\ea}{\end{eqnarray}}
\newcommand{\non}{\nonumber\\\rule{0pt}{30pt}}

\newcommand{\dis}{\displaystyle}
\newcommand{\eq}[1]{(\ref{#1})}

\newcommand{\Res}{\mathop{\rm Res}}
\newcommand{\rank}{\mathop{\rm rank}}

\newtheorem{lemma}{Lemma}[section]
\def\qed{\hfill\nobreak\hbox{$\square$}\par\medbreak}
\makeatletter
\@addtoreset{equation}{section}
\makeatother

\begin{document}

\begin{flushright}
LPENSL-TH-03/02\\
\end{flushright}

\par \vskip .1in \noindent
\vspace{24pt}
\begin{center}
\begin{LARGE}
{\bf Correlation functions of the $XXZ$
spin-$\textstyle{\frac{1}{2}}$ Heisenberg chain at the free fermion
point from their multiple integral representations}
\end{LARGE}
\vspace{50pt}
\begin{large}
{\bf N.~Kitanine}\footnote[1]{Graduate School of Mathematical
Sciences, University of Tokyo, Japan,
kitanine@ms.u-tokyo.ac.jp\par
\hspace{2mm} On leave of absence from Steklov Institute at
St. Petersburg, Russia},~~
{\bf J.~M.~Maillet}\footnote[2]{ Laboratoire de Physique, UMR 5672 du CNRS,
ENS Lyon,  France,
 maillet@ens-lyon.fr},~~
{\bf N.~A.~Slavnov}\footnote[3]{ Steklov Mathematical Institute,
Moscow, Russia, nslavnov@mi.ras.ru},~~
{\bf V.~Terras}\footnote[4]{NHETC, Department of Physics and Astronomy,
Rutgers University, USA, vterras@physics.rutgers.edu \par
\hspace{2mm} On leave of absence from LPMT, UMR 5825 du CNRS,
Montpellier, France}
\end{large}

\vspace{80pt}

\centerline{\bf Abstract}

\vspace{1cm}

\parbox{12cm}{\small%
Using multiple integral representations, we derive exact expressions
for the correlation functions of the
spin-$\textstyle{\frac{1}{2}}$ Heisenberg chain at the free fermion
point  $\Delta= 0$.}
\end{center}

\newpage

\section{Introduction}

In the article \cite{KitMST02} new multiple integral representations
for the correlation functions of the $XXZ$
spin-$\textstyle\frac{1}{2}$ Heisenberg chain have been obtained. In
the present article, we apply the results of \cite{KitMST02} to 
compute the correlation functions of the
spin-$\textstyle{\frac{1}{2}}$ Heisenberg chain at the free fermion
point  $\Delta= 0$. 

Generically, the Hamiltonian of the finite
cyclic $XXZ$ chain with $M$ sites (where $M$ is supposed to be even) has
the form
\be{0HamXXZ}
H_{\scriptscriptstyle XXZ}=\sum_{m=1}^{M}\left(
\sigma^x_{m}\sigma^x_{m+1}+\sigma^y_{m}\sigma^y_{m+1}
+\Delta(\sigma^z_{m}\sigma^z_{m+1}-1)
-\frac{h}{2}\sigma^z_{m}\right).
\ee
Here, $\sigma^{x,y,z}_{m}$ denote the local spin operators
(Pauli matrices) associated with the $m$-th site of the chain, $\Delta$
is the anisotropy parameter, and $h$ an external classical magnetic field.
The particularization of this model to the case $\Delta=0$ is known
as the $XX$ chain (isotropic $XY$ model \cite{LieSM61}):
\be{0HamXX}
H_{\scriptscriptstyle XX}=\sum_{m=1}^{M}\left(
\sigma^x_{m}\sigma^x_{m+1}+\sigma^y_{m}\sigma^y_{m+1}
-\frac{h}{2}\sigma^z_{m}\right).
\ee

In spite of the fact that the $XX$ spin-$\textstyle{\frac{1}{2}}$
chain is equivalent to a model of free fermions, its correlation functions 
are quite non-trivial. They had been studied for a long period
by numerous authors
\cite{Wu66,Mcc68,VadT78,MccPS83,ColIKT93,ItsIKS93} and the
key results in this field are presently known. It is worth mentioning
however that the methods used in the works listed above rely essentially
on the free fermion features of the $XX$ model. Therefore, they
cannot be applied to the more general $XXZ$ case, at
least without significant modifications. On the contrary, we expect
our present approach, which relies on multiple integral 
representations of correlation functions, 
to be instructive for the study of the general case as well. 

In 1992 \cite{JimMMN92},
multiple integral representations for the correlation functions of
the  $XXZ$ chain at zero temperature, $\Delta>1$ and $h=0$ have
been obtained from the $q$-vertex operator approach. Later, in
1996 \cite{JimM96} (see also \cite{JimML95}), similar answers were
formulated for the case $|\Delta|\le1$. A proof of these formulas,
together with their extension to non-zero magnetic field, has been
obtained in 1999 \cite{KitMT00,KitMT99} for both regimes using
algebraic Bethe ansatz and the actual resolution of the quantum
inverse scattering problem \cite{KitMT99,MaiT99}. It results from these
articles that, starting from the
so-called elementary blocks, one can in principle obtain 
a multiple integral representation for any
$n$-point correlation function of the $XXZ$ chain. 

More precisely, if $|\psi_g\rangle$ denotes the ground
state, and $E^{\epsilon'_m,\epsilon_m}_{m}$  the elementary
operators acting on the quantum space ${\cal H}_m$ at site $m$ as
the $2 \times 2$ matrices
$E^{\epsilon',\epsilon}_{lk}=\delta_{l,\epsilon'}\delta_{k,\epsilon}$,
the elementary blocks for correlation functions are defined as
\be{0genabcd}
F_m(\{\epsilon_j,\epsilon'_j\})=\frac
{\langle\psi_g|\prod\limits_{j=1}^m
E^{\epsilon'_j,\epsilon_j}_j|\psi_g\rangle}
{\langle\psi_g|\psi_g\rangle}.
\ee
The methods developed in \cite{JimMMN92}--\cite{KitMT00} enable
one to obtain the quantity \eq{0genabcd} in terms of
an integral with $m$ integrations.  It is clear that an arbitrary
correlation function in the ground state can be expressed as a
linear combination of the elementary blocks \eq{0genabcd} 
and, hence, as a linear combination of such multiple integrals. 
It should be stressed however that, 
although these formulas are quite explicit, the actual
analytic computation of these multiple integrals is missing up to
now. Moreover, the evaluation of correlations of physical
relevance, like spin-spin correlation functions, is a
priori quite involved. Indeed,  if we consider for example
the correlation function $\langle\sigma_1^z\sigma_{m}^z\rangle$,
the identity
\be{0szsz}
\langle\psi_g|\sigma^z_1\sigma^z_{m}|\psi_g\rangle
\equiv\langle\psi_g|(E^{11}_1-E^{22}_1)\prod\limits_{j=2}^{m-1}
(E^{11}_j + E^{22}_j) (E^{11}_{m}-E^{22}_{m})|\psi_g\rangle
\ee
shows that the corresponding linear combination of elementary blocks
is actually given as a sum of $2^{m}$ terms.
This means that the number of terms to sum up growths exponentially with $m$,
which in particular makes it extremely difficult to solve the problem
of asymptotic behavior at large distance.

Thus, up to recently, the situation in this field was as follows.
On one hand, the free fermion limit ($\Delta=0$) of the $XXZ$ model 
was well studied, but the extension of these results to the general case
came up against serious problems. On the other hand, the multiple integrals
approach formally provided the possibility to compute the
correlation functions for arbitrary $\Delta$; however, because
of the technical reasons mentioned above, no result
has up to now been reproduced via this method, even for the simplest case
of free fermions%
\footnote[1]{Recently, in \cite{ShiTN01},  
the probability to find in the ground
state a string of particles with spin down
(the emptiness formation probability) was computed at $\Delta=0$
by the method of multiple integrals. Let us stress that,
contrary to the spin-spin correlation functions, this quantity
can be expressed as a single elementary block. We consider
the emptiness formation probability in Section \ref{EPF} 
of the present article.}.

The  goal of this paper is to study the correlation functions of
the $XX$ chain using the new multiple integral representations obtained
in \cite{KitMST02}. In fact, these new representations are nothing
but re-summations of the multiple integral expressions for the elementary
blocks. In particular, they enable us to present the
spin-spin correlation functions of the type \eq{0szsz} as a sum of
only $m$ terms instead of $2^m$.
We would like to point out that in this paper we do not obtain new
results, but only reproduce the known answers via a new method. 
Moreover, we consider here the $XX$ model only as a test for the
relevance of the formulas obtained in \cite{KitMST02}. 
We hope that some of the
methods developed in the present publication can be applied (perhaps
after certain modifications) to the general $XXZ$ chain as
well.

This article is organized as follows. In the next section, we introduce
some useful notations and give the list of formulas obtained in
\cite{KitMST02} for the correlation functions of the $XXZ$ model. In
Section \ref{SZSZ}, we compute the correlation function of the third
components of spin. The emptiness formation probability is considered
in Section \ref{EPF}. In Section \ref{SS}, we obtain a Fredholm
determinant representation for the correlation function
$\langle\sigma_1^+\sigma_{m+1}^-\rangle$. The asymptotic analysis of
this Fredholm determinant is performed in Section \ref{LA}. Some
perspectives are discussed in the conclusion.


\section{Correlation functions of the $XXZ$ chain}\label{XXZ}

For the reader's convenience, we gather in this section the list of
results obtained in \cite{KitMST02} for the correlation functions of
the $XXZ$ model. 
Since eventually we study the limit $\Delta=0$,
we hereafter restrict ourselves to the case $|\Delta|<1$.

\medskip

Let us first of all recall that the Hamiltonian \eq{0HamXXZ} possesses
the symmetries
\be{XXZsym}
\begin{array}{ll}
{\dis
UH_{\scriptscriptstyle XXZ}(\Delta,h)U^{-1}=
-H_{\scriptscriptstyle XXZ}(-\Delta,h),}&
{\dis\qquad U=\prod_{m=1}^{M/2}\sigma_{2m}^z,}\non
{\dis
VH_{\scriptscriptstyle XXZ}(\Delta,h)V^{-1}=
H_{\scriptscriptstyle XXZ}(\Delta,-h),}&
{\dis\qquad V=\prod_{m=1}^{M}\sigma_{m}^x.}
\end{array}
\ee
Due to this freedom, there is no common definition of the $XXZ$ model.
It means that the expressions of correlation functions obtained in
different publications may coincide up to a common sign and/or the
sign of $\Delta$ and $h$.

The standard parameterization of the anisotropy parameter is
$\Delta=\cosh\eta$. In the regime $|\Delta|<1$ the parameter $\eta$
is  imaginary, and we set $\eta=-i\zeta$, $\zeta>0$.
The free fermion point $\Delta=0$ corresponds to $\zeta=\pi/2$.

\medskip

The general structure of the expressions obtained in \cite{KitMST02} 
for the spin-spin correlation
functions is the following:
\be{XXZnewGF}
\langle\sigma^{\alpha}_1\sigma_{m+1}^{\beta}\rangle=
\sum_{n=0}^{m-1}\ \oint\limits_{C_z}d^{n+1} z
\int\limits_{C_{\lambda}} d^{n}\lambda \int\limits_{C_{\mu}}
d^{2}\mu\ f^{m}(\{\lambda,z\})\ \Gamma^{\alpha
\beta}_n(\{\lambda,\mu,z\})\ S_h(\{\lambda,z\}).
\ee
Here $\alpha,\beta = x,y,z$, and the functions  $\Gamma^{\alpha
\beta}_n(\{\lambda,\mu,z\})$ and $f(\{\lambda,z\})$ are purely
algebraic quantities, which in particular do not depend on the regime nor
on the magnetic field. Their explicit forms for specific correlation
functions are given below.

The integration contour $C_z$ surrounds the point $z_j=-i\zeta/2$
($z_j=-i\pi/4$ for $\Delta=0$), where the function
$f(\{\lambda,z\})$ has a pole. All other singularities of the
integrand \eq{XXZnewGF} are outside the contour $C_z$.

The contours $C_\lambda$ and $C_\mu$ depend on $\Delta$ and $h$.
In all the examples considered below we have $C_\lambda=C_\mu$.
For $|\Delta|<1$ and $h\ge0$, the contour $C_\lambda$ is an interval
$[-\Lambda,\Lambda]$ of the real axis, where the value of $\Lambda$ is
uniquely defined by $\Delta$ and $h$, although in the general case the
dependency $\Lambda=\Lambda(\Delta,h)$ is rather implicit. At
$\Delta=0$, however, the integration domain can be found explicitly from
the fact that $\cosh2\Lambda =4/h$. 
Note that if $h\to 0$, one has $\Lambda\to\infty$. On the
other hand, if $h$ approaches its critical value $h_c=4$, then
$\Lambda\to0$ and all the correlation functions become trivial,
which comes physically from the fact that the ground state of the Hamiltonian
\eq{0HamXX} is then purely ferromagnetic. Therefore we consider below
only the case $0\le h\le h_c$. Due to the symmetry \eq{XXZsym}, we
also do not need to consider negative magnetic field, although all
our results remain valid for $h<0$ as well\footnote[1]{%
Actually the restriction $h\ge0$ is not necessary. In the case
$h<0$, the integration contour $C_\lambda$ becomes\\
$[-\Lambda+\frac{i\pi}{2},-\infty+\frac{i\pi}{2}]\cup
\mathbb{R}\cup
[+\infty+\frac{i\pi}{2},\Lambda+\frac{i\pi}{2}]$, where 
$\Lambda$ is the real positive solution of the equation 
$\cosh2\Lambda=4/|h|$.}.

Finally, the integrand  \eq{XXZnewGF} contains a function
$S_h(\{\lambda,z\})$, which also depends on the value of the magnetic
field. This function is equal to the determinant of a matrix
of elements $\rho(\lambda_j,z_k)$,
where $\rho(\lambda,z)$ is the so-called `inhomogeneous density',
solution of the integral equation
\be{XXZinteqinh}
-2\pi i\rho(\lambda,z) +\int\limits_{-\Lambda}^\Lambda
K(\lambda-\mu)\rho(\mu,z)\,d\mu= t(\lambda,z),
\ee
with
\be{XXZKkern}
K(\lambda)=\frac{i\sin2\zeta}
{\sinh(\lambda+i\zeta)\sinh(\lambda-i\zeta)},
\qquad
t(\lambda,z)=\frac{-i\sin\zeta}
{\sinh(\lambda-z)\sinh(\lambda-z-i\zeta)}.
\ee
Note that, at $z=-i\zeta/2$, the function $\rho(\lambda,z)$ coincides
with the spectral density of the ground state. In the
free fermion limit $\Delta=0$ ($\zeta=\pi/2$), one has
$K(\lambda)=0$, and thus
\be{XXZrho}
\rho(\lambda,z)=\frac{i}{2\pi}t(\lambda,z)=
\frac{i}{\pi\sinh2(\lambda-z)}.
\ee
 
\medskip

After this general setting, let us now be more specific and  
present the explicit formulas for some of the correlation functions 
of the $XXZ$ chain in the domain $|\Delta|<1$.
We consider below essentially three different cases:
 
a) the correlation function 
$g_m^{zz}=\langle\sigma_1^z\sigma_{m+1}^z\rangle$ in a
magnetic field;
 
b) the emptiness formation probability $\tau(m)$ in a
magnetic field;
 
c) the correlation functions
$g_m^{+-}=\langle\sigma_1^+\sigma_{m+1}^-\rangle$ in zero
magnetic field. 

\par\noindent 
The reader can find the derivation of the
corresponding multiple integral representations in \cite{KitMST02}.

\medskip

a) The correlation function of the third components of spin
can be evaluated from the generating functional
$\langle\exp(\beta Q_{1,m})\rangle$,
where $Q_{1,m}=\frac12\sum_{k=1}^{m}(1-\sigma_k^z)$, as 
\be{SZgenfun}
\langle\sigma_1^z\sigma_{m+1}^z\rangle=
\left(2{\cal D}^2_m\left.\frac{
\partial^2}{\partial\beta^2}
-4{\cal D}_m\frac{
\partial}{\partial\beta}+1\right)
\langle\exp(\beta Q_{1,m})\rangle
\right|_{\beta=0}=
2{\cal D}^2_m\langle Q^2_{1,m}\rangle
-4{\cal D}_m\langle Q_{1,m}\rangle+1.
\ee
Here, the symbols ${\cal D}_m$ and ${\cal D}^2_m$ denote respectively 
the first and the second lattice derivative,
\be{SZlatder}
{\cal D}_mf(m)\equiv f(m+1)-f(m),\qquad
{\cal D}_m^2f(m)\equiv f(m+1)+f(m-1)-2f(m).
\ee
The expectation value of the functional
$\langle\exp(\beta Q_{1,m})\rangle$ is given by (5.8) of
\cite{KitMST02}:
\ba{SZGFres}
&&{\dis\hspace{-2mm}
\langle\exp(\beta Q_{1,m})\rangle=
\sum_{n=0}^m\frac1{(n!)^2}
\oint\limits_{C_z}
\prod_{j=1}^n\frac{dz_j}{2\pi i}
\int\limits_{-\Lambda}^\Lambda d^{n}\lambda \cdot
\prod_{a=1}^n \left(\frac{\sinh(z_a-\frac{i\zeta}2)
\sinh(\lambda_a+\frac{i\zeta}2)}
{\sinh(z_a+\frac{i\zeta}2)\sinh(\lambda_a-\frac{i\zeta}2)}\right)^{m}
}\non
&&{\dis\hspace{8mm}
\times
W_n(\{\lambda\}|\{z\})\cdot
{\det}_{n}\Bigl[\tilde M_{jk}(\{\lambda\}|\{z\})\Bigr]
\cdot{\det}_{n}\Bigl[\rho(\lambda_j,z_k)\Bigr].}
\ea
Here and further we use the notation ${\det}_n$ for the determinants
of $n\times n$ matrices. The function $W_n$ is defined by
\be{SZWn}
W_n(\{\lambda\},\{z\})=
\prod_{a=1}^n\prod_{b=1}^n\frac{
\sinh(\lambda_a-z_b-i\zeta)\sinh(z_b-\lambda_a-i\zeta)}
{\sinh(\lambda_a-\lambda_b-i\zeta)\sinh(z_a-z_b-i\zeta)},
\ee
and the  entries of the matrix $\tilde M_{jk}$ are 
\be{SZtiMjk}
\hspace{-2mm}
\tilde M_{jk}(\{\lambda\}|\{z\})
=t(z_k,\lambda_j)+e^\beta t(\lambda_j,z_k)
\prod_{a=1}^n\frac{\sinh(\lambda_a-\lambda_j-i\zeta)}
{\sinh(\lambda_j-\lambda_a-i\zeta)}
\frac{\sinh(\lambda_j-z_a-i\zeta)}
{\sinh(z_a-\lambda_j-i\zeta)},
\ee
where the functions $t(\lambda,z)$ and $\rho(\lambda,z)$ are
defined in \eq{XXZKkern}  and \eq{XXZinteqinh}  respectively.

For $h=0$, it is more convenient to derive the correlation function
$\langle\sigma_{1}^z\sigma_{m+1}^z\rangle$ from  the generating
functional $\langle\exp(\beta Q_{1,m})\sigma_{m+1}^z\rangle$:
\be{XZZszsz}
\langle\sigma_1^z\sigma_{m+1}^z\rangle=
\left.-2{\cal D}_{m-1}\frac{\partial}{\partial\beta}
\langle\exp(\beta Q_{1,m})\sigma_{m+1}^z\rangle
\right|_{\beta=0},\qquad h=0.
\ee
The expectation value of the functional
$\langle\exp(\beta Q_{1,m})\sigma_{m+1}^z\rangle$ is given by
(6.7) of \cite{KitMST02}:
\ba{XXZGF}
&&{\dis\hspace{-2mm}
\langle\exp(\beta Q_{1,m})\sigma_{m+1}^z\rangle=
\sum_{n=0}^m\frac1{(n!)^2}
\oint\limits_{C_z}
\prod_{j=1}^n\frac{dz_j}{2\pi i}
\int\limits_{\mathbb{R}} d^{n+1}\lambda \cdot
\prod_{a=1}^n \left(\frac{\sinh(z_a-\frac{i\zeta}2)
\sinh(\lambda_a+\frac{i\zeta}2)}
{\sinh(z_a+\frac{i\zeta}2)
\sinh(\lambda_a-\frac{i\zeta}2)}\right)^{m}
}\non
&&{\hspace{-2mm} \times
\prod_{a=1}^n \left(\frac{\sinh(\lambda_a+\frac{i\zeta}2)}
{\sinh(z_a+\frac{i\zeta}2)}\right)
\left(\prod_{a=1}^n \frac{\sinh(\lambda_{n+1}-z_a)}
{\sinh(\lambda_{n+1}-\lambda_a)} - \prod_{a=1}^n
\frac{\sinh(\lambda_{n+1}-z_a-i\zeta)}
{\sinh(\lambda_{n+1}-\lambda_a-i\zeta)}\right)\cdot
W_n(\{\lambda\}|\{z\})}\non
&&{\dis\hspace{8mm}
\times
{\det}_{n}\Bigl[\tilde M_{jk}(\{\lambda\}|\{z\})\Bigr]
{\det}_{n+1}\Bigl[\rho(\lambda_j,z_1),\dots,
\rho(\lambda_j,z_n),
\rho(\lambda_j,{\textstyle-\frac{i\zeta}{2}})\Bigr].}
\ea

\medskip

b) The emptiness formation probability  (the probability to find in the
ground state a string of particles with spin down in the first $m$
sites) is defined as
\be{EMPtau}
\tau(m)=\langle\prod_{k=1}^m\frac{1-\sigma_k^z}2\rangle.
\ee
The multiple integral representation of this correlation
function can be easily obtained from the generating functional
$\langle\exp(\beta Q_{1,m})\rangle$
(see equation (C.9) of \cite{KitMST02}):
\be{EFPtausym}
\tau(m)=\lim_{\xi_1,\dots,\xi_m\to-\frac{i\zeta}{2}}
\frac1{m!}\int\limits_{-\Lambda}^{\Lambda}
\frac{Z_m(\{\lambda\},\{\xi\})}
{\prod\limits_{a<b}^m
\sinh(\xi_a-\xi_b)}
{\det}_m~\rho(\lambda_j,\xi_k)\,d^m\lambda,
\ee
where
\be{EFPZm}
Z_m(\{\lambda\},\{\xi\})=
\prod\limits_{a=1}^{m}\prod\limits_{b=1}^{m}
\frac{\sinh(\lambda_a-\xi_b)\sinh(\lambda_a-\xi_b-i\zeta)}
{\sinh(\lambda_a-\lambda_b-i\zeta)}
\cdot\frac{{\det}_m~t(\lambda_j,\xi_k)}{
\prod\limits_{a>b}^m\sinh(\xi_a-\xi_b)}.
\ee

\medskip

c) Let us consider finally the correlation functions
$g_m^{+-}=\langle\sigma_1^+\sigma_{m+1}^-\rangle$ and
$g_m^{-+}=\langle\sigma_1^-\sigma_{m+1}^+\rangle$. It is easy
to see that $g_m^{+-}(h)=g_m^{-+}(-h)$. For zero magnetic field, these
two correlation functions coincide. In this case their multiple
integral representation is given by (6.13) of \cite{KitMST02}:
\ba{SSs+s-}
&&{\dis\hspace{-7mm}
\langle \sigma^+_1\sigma^-_{m+1}\rangle=\sum_{n=0}^{m-1}
\frac1{n!(n+1)!}\oint\limits_{C_z}
 \prod_{j=1}^{n+1}\frac{dz_j}{2\pi i}
\int\limits_{\mathbb{R}} d^{n+2}\lambda
\prod_{a=1}^{n+1} \left(\frac{\sinh(z_a-\frac{i\zeta}2)}
{\sinh(z_a+\frac{i\zeta}2)}
\right)^m \hspace{2mm}
\prod_{a=1}^{n}\left(\frac{\sinh(\lambda_a+\frac{i\zeta}2)}
{\sinh(\lambda_a-\frac{i\zeta}2)}\right)^{m}
}\non
&&{\dis\hspace{7mm}
\times \frac1{\sinh(\lambda_{n+1}-\lambda_{n+2})}\cdot
\left(\frac{\prod\limits_{a=1}^{n+1}
\sinh(\lambda_{n+1}-z_a-{i\zeta})\sinh(\lambda_{n+2}-z_a)}
{\prod\limits_{a=1}^n\sinh(\lambda_{n+1}-\lambda_a-{i\zeta})
\sinh(\lambda_{n+2}-\lambda_a)}
\right)}\non
&&{\dis\hspace{7mm}
\times \hat W_n(\{\lambda\},\{z\})\cdot {\det}_{n+1} \hat M_{jk}\cdot
{\det}_{n+2}\left[
\rho(\lambda_j,z_1),\dots,\rho(\lambda_j,z_{n+1}),
\rho(\lambda_j,{-\textstyle\frac{i\zeta}{2}})
\right].}
\ea
Here,
\be{SShWn}
\hat W_n(\{\lambda\},\{z\})=
\frac{\prod\limits_{a=1}^n\prod\limits_{b=1}^{n+1}
\sinh(\lambda_a-z_b-i\zeta)\sinh(z_b-\lambda_a-i\zeta)}
{\prod\limits_{a=1}^n\prod\limits_{b=1}^n
\sinh(\lambda_a-\lambda_b-i\zeta)
\prod\limits_{a=1}^{n+1}\prod\limits_{b=1}^{n+1}
\sinh(z_a-z_b-i\zeta)},
\ee
and the entries of the $(n+1)\times(n+1)$ matrix $\hat M$ are
\be{SShM}
\begin{array}{ll}
  {\dis \hat M_{jk}=t(z_k,\lambda_j)-t(\lambda_j,z_k)
\prod\limits_{a=1}^n\frac{\sinh(\lambda_a-\lambda_j-i\zeta)}
{\sinh(\lambda_j-\lambda_a-i\zeta)}
\prod\limits_{b=1}^{n+1}\frac{\sinh(\lambda_j-z_b-i\zeta)}
{\sinh(z_b-\lambda_j-i\zeta)}},\  & j\le n,\\
  {\dis\hat M_{n+1,k}=t(z_k,{-\textstyle \frac{i\zeta}2})},  & j=n+1.
\end{array}
\ee

\medskip

To conclude this section, let us recall once more that all the
multiple integral representations given above hold for arbitrary
$-1<\Delta<1$. In order to particularize these expressions to the
case of the $XX$ model, one has to set $\zeta=\pi/2$, 
$\Lambda=\mathrm{arccosh} (4/h)/2$, and to use the expression \eq{XXZrho} for the
inhomogeneous density $\rho(\lambda,z)$.


\section{Correlation function of the third components of spin}\label{SZSZ}

Let us begin our calculations with the correlation function
$\langle\sigma_1^z\sigma_{m+1}^z\rangle$, which is the simplest
spin-spin correlation function of the $XX$ model.

Observe that, at $\zeta=\pi/2$, the equation \eq{SZGFres}
simplifies drastically.  First of all, the matrix $\tilde M_{jk}$
\eq{SZtiMjk} becomes proportional to the Cauchy matrix
\be{SZMjkfree}
\tilde M_{jk}(\{\lambda\}|\{z\})=
\frac{2(e^\beta-1)}{\sinh2(\lambda_j-z_k)},\qquad \zeta=\frac\pi2,
\ee
and, hence, to compute its determinant one can use the identity
\be{SZCauchy}
{\det}_n\frac1{\sinh(x_j-y_k)}=
\frac{\prod\limits_{j>k}^n\sinh(x_j-x_k)\sinh(y_k-y_j)}
{\prod\limits_{j,k=1}^n\sinh(x_j-y_k)}.
\ee
However, it is more important to notice that
${\det}_n\tilde M_{jk}$ is proportional to
$(e^\beta-1)^n$: this means that, if one takes the first 
(respectively the second) derivative with
respect to $\beta$ and sets $\beta=0$ as in \eq{SZgenfun}, 
only the terms $n\le 1$ (respectively $n\le2$) in
the sum \eq{SZGFres} do not vanish. 
Thus, after some simple computation, one obtains
\be{SZ1betader}
\langle Q_{1,m}\rangle=\frac1{4\pi^2}
\oint\limits_{C_z}
dz \int\limits_{-\Lambda}^{\Lambda}
\varphi^m(z)\varphi^{-m}(\lambda)
\frac{d\lambda }{\sinh^2(\lambda-z)},
\ee
\be{SZ2betader}
\langle Q^2_{1,m}\rangle=
\langle Q_{1,m}\rangle+
\frac1{32\pi^4}
\oint\limits_{C_z}
d^2z \int\limits_{-\Lambda}^{\Lambda}
d^{2}\lambda
\prod_{a=1}^2 \Bigl(\varphi^m(z_a)\varphi^{-m}(\lambda_a)\Bigr)
\left({\det}_2\frac1{\sinh(\lambda_j-z_k)}\right)^2,
\ee
where we have introduced the notation
\be{SZphi}
\varphi(z)=\frac{\sinh(z-\frac{i\pi}4)}{\sinh(z+\frac{i\pi}4)}.
\ee

Let us first consider the integral \eq{SZ1betader}. 
As the contour $C_z$ surrounds only the singularity $z=-i\pi/4$, 
where $\varphi(z)$ admits a pole of order $m$, the value of the $z$-integral 
in \eq{SZ1betader} is given by the corresponding residue at $z=-i\pi/4$. 
However, this way to compute the $z$-integral is not very convenient, 
especially for large $m$. 
Instead, we suggest to deform the original contour $C_z$ into
an infinite horizontal strip of boundary $\Gamma$ given by $\Im
z=z_0-\pi$ and $\Im z=z_0$, where $0<z_0<3\pi/4$.  Then, obviously,
\be{SZstrip}
\oint\limits_{C_z}
\frac{\varphi^m(z)\,dz}{\sinh^2(\lambda-z)}=
\oint\limits_{\Gamma}
\frac{\varphi^m(z)\,dz}{\sinh^2(\lambda-z)}
-2\pi i\Res\left.
\frac{\varphi^m(z)}{\sinh^2(\lambda-z)}\right|_{z=\lambda}.
\ee
Since the integrand is a periodic
function with period $i\pi$, and since it vanishes at $z\to\pm\infty$,
it is clear that the integral with respect to the new contour 
$\Gamma$ is equal to zero.
Thus, to compute the $z$-integral in \eq{SZ1betader}, it is enough to take
the residue in the second order pole at $z=\lambda$.
The remaining integral with respect
to $\lambda$ is then trivially computable, and we obtain
\be{SZQ0}
\langle Q_{1,m}\rangle=\frac{m}\pi\arctan(\sinh2\Lambda).
\ee
In the next sections, we shall deal with the change of integration
variables $\cosh2\lambda= (\cos p)^{-1}$. Therefore, let us at this stage 
introduce $p_0$ such that $\cosh2\Lambda= (\cos p_0)^{-1}$. 
Then \eq{SZQ0} takes the form
\be{SZQ}
\langle Q_{1,m}\rangle=\frac{mp_0}\pi,
\ee
where $p_0=\arccos\left(\frac{h}{4}\right)$.

The integral \eq{SZ2betader} can be taken in the same manner. The
integration with respect to $z_1$ and $z_2$ leads to
\be{SZQ2}
\langle Q_{1,m}^2\rangle=
\frac{mp_0}\pi+\left(\frac{mp_0}\pi\right)^2
+\frac1{4\pi^2}\int\limits_{-\Lambda}^{\Lambda}
\left(\frac{\varphi^{\frac m2}(\lambda_1)
\varphi^{-\frac m2}(\lambda_2)-
\varphi^{-\frac m2}(\lambda_1)\varphi^{\frac m2}(\lambda_2)}
{\sinh(\lambda_1-\lambda_2)}\right)^2\,d\lambda_1\,d\lambda_2.
\ee
In fact, we do not need to compute $\langle Q_{1,m}^2\rangle$
itself, but only its second lattice derivative.
Differentiating \eq{SZQ2} with respect to $m$, we immediately
arrive at
\be{SZD_mQ2}
{\cal D}^2_m\langle Q_{1,m}^2\rangle=
2\left(\frac{p_0}\pi\right)^2-\frac1{\pi^2m^2}
(1-\cos2mp_0).
\ee

Combining \eq{SZQ} and \eq{SZD_mQ2}, we finally obtain
\be{SZszsz}
\langle\sigma_1^z\sigma_{m+1}^z\rangle=
\left(\frac{2p_0}\pi-1\right)^2-\frac2{\pi^2m^2}
(1-\cos2mp_0).
\ee

It is worth mentioning that, in spite of the fact that we have
formally restricted ourselves to the
case $h\ge0$, the result \eq{SZszsz} remains valid for $h<0$ as well.
For zero magnetic field, $p_0=\pi/2$, and the constant contribution
to the correlation function disappears. In order to get rid of this
constant term from the very beginning, it is more convenient to derive
$\langle\sigma_1^z\sigma_{m+1}^z\rangle$ from  the generating
functional $\langle\exp(\beta Q_{1,m})\sigma_{m+1}^z\rangle$
(see \eq{XXZGF}). Then, for $\Delta=0$, the corresponding sum 
reduces to the single term $n=1$, which gives
\be{SZanothint}
\langle\sigma_1^z\sigma_{m+1}^z\rangle
=\frac{2i}{\pi^3}
\oint\limits_{C_z}\,dz
\int\limits_{\mathbb{R}}\frac{d\lambda_1d\lambda_2}
{\cosh2\lambda_2\cosh2\lambda_1}\cdot
\frac{\varphi^{m}(z)\varphi^{-m}(\lambda_1) }
{\sinh2(\lambda_2-z)}.
\ee
Using the method of calculations described
above, we find
\be{SZszsz1}
\langle\sigma_1^z\sigma_{m+1}^z\rangle=
\frac2{\pi^2m^2}\Bigl((-1)^m-1\Bigr),\qquad \mbox{at}\quad h=0.
\ee
%


\section{Emptiness formation probability}\label{EPF}

The emptiness formation probability \eq{EMPtau} constitute 
one of the simplest example of correlation functions. In particular, 
unlike the spin-spin correlation functions studied in Sections \ref{SZSZ},
\ref{SS} and \ref{LA},
it can be directly expressed as a single elementary block of the form 
\eq{0genabcd}, therefore as a single (multiple) integral which can be
written in the symmetric form \eq{EFPtausym}. In this section,
we explain how to compute this integral in the case $\Delta=0$, 
and how to analyze its asymptotic
behavior in the limit $m\to\infty$.
 
Setting $\zeta=\pi/2$ in the equations \eq{EFPtausym},  \eq{EFPZm},
we have
\be{EFPff}
\tau(m)=\lim_{\xi_1,\dots,\xi_m\to-\frac{i\pi}{4}}
\frac1{m!}\int\limits_{-\Lambda}^{\Lambda}
\frac{Z_m(\{\lambda\},\{\xi\})}
{\prod\limits_{a<b}^m
\sinh(\xi_a-\xi_b)}
{\det}_m\left(\frac{i}{\pi\sinh2(\lambda_j-\xi_k)}\right)
\,d^m\lambda,
\ee
and
\be{EFPZmff}
Z_m(\{\lambda\},\{\xi\})=
2^{-m^2}\prod\limits_{a=1}^{m}\prod\limits_{b=1}^{m}
\frac{\sinh2(\lambda_a-\xi_b)}
{\cosh(\lambda_a-\lambda_b)}
\prod\limits_{a>b}^m\sinh^{-1}(\xi_a-\xi_b)
{\det}_m\left(\frac{2}{\sinh2(\lambda_j-\xi_k)}\right).
\ee
Once again, we have to deal with determinants of Cauchy matrices.
Computing them via \eq{SZCauchy} and setting $\xi_j=-i\pi/4$, we obtain
\be{EFPtau-m-ff}
\tau(m)=\frac{2^{m^2}}{m!(2\pi)^m}
\int\limits_{-\Lambda}^{\Lambda}
\frac{\prod\limits_{a>b}^m
\sinh^2(\lambda_a-\lambda_b)}
{\prod\limits_{a=1}^m
\cosh^m(2\lambda_a)}\,d^m\lambda.
\ee
The representation \eq{EFPtau-m-ff} can be reduced to a Toeplitz
determinant. Indeed, the change of variables
$\cosh2\lambda_j=\cos^{-1} p_j$ leads to
\be{EMPtau-m-p}
\tau(m)=\frac{2^{m^2-m}}{m!(2\pi)^m}
\int\limits_{-p_0}^{p_0}
\prod\limits_{a>b}^m
\sin^2\frac{(p_a-p_b)}2\,d^mp=
\frac{1}{m!(2\pi)^m}
\int\limits_{-p_0}^{p_0}
\Delta(e^{-ip})\Delta(e^{ip})\,d^mp,
\ee
where  $\Delta(e^{\pm ip})$ denote Van-der-Monde determinants of
variables $e^{\pm ip_j}$. Due to the symmetry of the integrand with
respect to all $p_j$, one can replace one of these Van-der-Monde
determinants with the product of its diagonal elements multiplied by
$m!$, which gives us
\be{EMPToeplitz}
\tau(m)=\frac{1}{(2\pi)^m}
\int\limits_{-p_0}^{p_0}
\prod_{k=1}^{m}e^{-i(k-1)p_k}
{\det}_m\left(e^{i(j-1)p_k}\right)\,d^mp
={\det}_m\left(\frac{1}{2\pi}
\int\limits_{-p_0}^{p_0}
e^{i(j-k)p}\,dp\right).
\ee
The representation \eq{EMPToeplitz} of $\tau(m)$ as a Toeplitz determinant 
has already been obtained in
\cite{ShiTN01} from the multiple integral representation given in
\cite{KitMT00}.

Thus, \eq{EMPToeplitz} provides an explicit expression 
of the emptiness formation probability, at least if $m$ is small enough. 
However, it is more
important to be able to extract the asymptotic behavior of $\tau(m)$ at
$m\to\infty$. There exist several ways to do this. Firstly, one can
analyze the determinant \eq{EMPToeplitz} as in \cite{ShiTN01}.
Secondly, the determinant \eq{EMPToeplitz} can be transformed to
a Fredholm determinant of a linear integral operator
\cite{ColIKT93}, the asymptotic behavior of which can be
evaluated from the matrix Riemann-Hilbert problem \cite{DIZ}.  Here
we propose a third approach, based on the application of the saddle
point method directly to \eq{EFPtau-m-ff}. It is possible that
this method can be used also for the general $XXZ$ model.

Let us rewrite \eq{EFPtau-m-ff} in the following way:
\be{EMPstart}
\tau(m)=\frac{2^{m^2}}{(2\pi)^m}\int_D
e^{m^2S(\{\lambda\})}\,d\lambda,
\ee
where
\be{EMPaction}
S(\{\lambda\})=
\frac{1}{m^2}\sum_{a>b}^m
\log\sinh^2(\lambda_a-\lambda_b)
-\frac{1}{m}\sum_{a=1}^m
\log\cosh2\lambda_a,
\ee
and the domain $D$ is defined by $-\Lambda\le\lambda_1\le\dots\le\lambda_m
\le\Lambda$. The integrand in \eq{EMPstart} is positive within the domain
$D$ and vanishes on its boundary. Moreover, it is not difficult to
check that the matrix of the second derivatives
$\partial^2S/\partial\lambda_j\partial \lambda_k$ is negatively
defined. Hence, the integrand has a unique maximum in $D$, which is given
by the system
\be{EMPsystem}
m\frac{\partial S}{\partial\lambda_j}=
\frac{2}{m}\sum_{a=1\atop{a\ne j}}^m\coth(\lambda_j-\lambda_a)
-2\tanh2\lambda_j=0.
\ee
Following the standard arguments of the saddle point method, we assume
that, in the limit $m\to\infty$, the solutions of the system
\eq{EMPsystem} are distributed on the interval $[-\Lambda, \Lambda]$ 
according to a certain density $\rho_0(\lambda)$. 
Then, in this limit, \eq{EMPsystem} becomes
an integral equation for this density:
\be{EMPinteql}
\tanh2\lambda=
V.P.\int\limits_{-\Lambda}^{\Lambda}\coth(\lambda-\mu)
\rho_0(\mu)\,d\mu.
\ee
In its turn, the integral \eq{EMPstart} can be
approximated by the value of the integrand in the saddle point:
\be{EMPasym}
\tau(m)\to 2^{m^2}e^{m^2S_0}, \qquad m\to \infty,
\ee
where
\be{EMPS0}
S_0=
\frac{1}{2}\int\limits_{-\Lambda}^{\Lambda}
\log\sinh^2(\lambda-\mu)\rho_0(\lambda)\rho_0(\mu)\,d\lambda d\mu
-\int\limits_{-\Lambda}^{\Lambda}\log\cosh2\lambda\cdot
\rho_0(\lambda)\,d\lambda.
\ee
The analytic expression of the function $\rho_0(\lambda)$ can be 
determined as follows. Setting
$x=e^{2\lambda}$, we transform \eq{EMPinteql} into
\be{EMPint-eq}
\frac{x}{x^2+1}=V.P.\int_a^b\frac{dy}{x-y}\hat\rho_0(y),
\ee
where $\hat\rho_0(x)=\frac{1}{2}e^{-2\lambda}\rho_0(\lambda)$ and
$a=e^{-2\Lambda},\quad b=e^{2\Lambda}$.
The solution of the singular integral equation \eq{EMPint-eq}
can be obtained in a standard way via the scalar Riemann--Hilbert problem.
Let us define
\be{EMPf}
f_\pm(x)=\int_a^b\frac{dy}{x-y\pm i0}\hat\rho_0(y).
\ee
Then we have
\be{EMPeqrho}
2\pi i\hat\rho_0(x)=f_-(x)-f_+(x),\qquad
\int_a^b\hat\rho_0(x)\,dx=1.
\ee
At the same time, $f(x)$ satisfies the relation
\be{EMPRHP}
f_-(x)+f_+(x)=\frac{2x}{x^2+1},
\ee
and, hence,
\be{EMPsolRHP}
f(x)=f^0(x)\left\{
C+\frac{1}{\pi i}\int_a^b\frac{dy}{y-x}\cdot
\frac{y}{(y^2+1)f_+^0(y)}\right\},
\ee
where $f^0(x)=\Bigl((x-a)(x-b)\Bigr)^{-1/2}$ and $C$ is a constant.
Substituting this expression into \eq{EMPeqrho}, we eventually obtain
\be{EMPanshrho}
\hat\rho_0(x)=\frac{1}{\pi}\frac{x+1}{x^2+1}
\sqrt{\frac{a+b}{2(x-a)(b-x)}},
\ee
which results into
\be{EMPansrho}
\rho_0(\lambda)=\frac{1}{\pi}\frac{\cosh\lambda}{\cosh2\lambda}
\sqrt{\frac{\cosh2\Lambda}
{\sinh(\Lambda-\lambda)\sinh(\Lambda+\lambda)}}.
\ee

This enables us to obtain the analytic expression of $S_0$ in terms of 
$\Lambda$:
\be{EMPansS0}
S_0=-\log\left(2\frac{\sqrt{\cosh2\Lambda}}{\sinh\Lambda}
\right).
\ee
Taking into account that $\cosh2\Lambda=4/h$, we finally
obtain the following asymptotic equivalent of $\tau(m)$ in terms of the
magnetic field $h$:
\be{EMPansasym}
\tau(m)\to \left(\frac{4-h}{8}\right)^{\frac{m^2}{2}},
\qquad m\to \infty.
\ee

Thus, this method provides an alternative derivation of 
the asymptotic behavior of the emptiness formation probability \cite{DIZ}.


\section{Correlation function $\langle \sigma^+_1\sigma^-_{m+1}\rangle$ 
as Fredholm determinant}\label{SS}

Unlike the correlations of the third components of spin 
(see Section \ref{SZSZ}),
the correlation function $\langle \sigma^+_1\sigma^-_{m+1}\rangle$
remains non-trivial even at the free fermion point $\Delta=0$.  
In this section,
we show how to compute it from \eq{SSs+s-} for zero magnetic field.

Let us first consider the part of the integrand \eq{SSs+s-} which depends on
$\lambda_{n+1}$ and $\lambda_{n+2}$:
$$
\frac{{\det}_{n+2}~\rho(\lambda_j,z_k)}
{\sinh(\lambda_{n+1}-\lambda_{n+2})}\cdot
\left(\frac{\prod\limits_{a=1}^{n+1}
\sinh(\lambda_{n+1}-z_a-{i\zeta})\sinh(\lambda_{n+2}-z_a)}
{\prod\limits_{a=1}^n\sinh(\lambda_{n+1}-\lambda_a-i\zeta)
\sinh(\lambda_{n+2}-\lambda_a)}
\right),
$$
where one should set $z_{n+2}\equiv -i\zeta/2$ in the last column of
the determinant of densities. We can shift the
integration contour for $\lambda_{n+2}$ by $-i\zeta$, and then
replace $\lambda_{n+2}$ by $\lambda_{n+2}-i\zeta$ in the integrand.  This
changes the sign of the density function, and we obtain
$$
-\frac{{\det}_{n+2}~\rho(\lambda_j,z_k)}
{\sinh(\lambda_{n+1}-\lambda_{n+2}+i\zeta)}\cdot
\left(\frac{\prod\limits_{a=1}^{n+1}
\sinh(\lambda_{n+1}-z_a-{i\zeta})\sinh(\lambda_{n+2}-z_a-i\zeta)}
{\prod\limits_{a=1}^n\sinh(\lambda_{n+1}-\lambda_a-{i\zeta})
\sinh(\lambda_{n+2}-\lambda_a-i\zeta)}
\right).
$$
We see that for $\zeta=\pi/2$ the integrand becomes an
antisymmetric function of $\lambda_{n+2}$ and $\lambda_{n+1}$ and, hence,
the corresponding integral vanishes. It remains then to take into account
the contribution of the poles which have been crossed during the shift of
the $\lambda_{n+2}$-contour.  It is easy to see that we have
crossed only one singularity, which corresponds to the pole of the function
$\rho(\lambda_{n+2},{-\textstyle\frac{i\zeta}{2}})$ at
$\lambda_{n+2}={-\textstyle\frac{i\zeta}{2}}$.  The residue in this
point gives
\ba{SSs+s-new}
&&{\dis\hspace{-7mm}
\langle \sigma^+_1\sigma^-_{m+1}\rangle=\sum_{n=0}^{m-1}
\frac{1}{n!(n+1)!}\oint\limits_{C_z}
 \prod_{j=1}^{n+1}\frac{dz_j}{2\pi i}
\int\limits_{\mathbb{R}} d^{n+1}\lambda
\prod_{a=1}^{n+1} \left(\frac{\sinh(z_a-\frac{i\zeta}2)}
{\sinh(z_a+\frac{i\zeta}2)}
\right)^m \hspace{2mm}
\prod_{a=1}^{n}\left(\frac{\sinh(\lambda_a+\frac{i\zeta}2)}
{\sinh(\lambda_a-\frac{i\zeta}2)}\right)^{m}
}\non
&&{\dis\hspace{27mm}
\times \frac1{\sinh(\lambda_{n+1}+\frac{i\zeta}2)}\cdot
\left(\frac{\prod\limits_{a=1}^{n+1}
\sinh(\lambda_{n+1}-z_a-{i\zeta})\sinh(z_a+\frac{i\zeta}2)}
{\prod\limits_{a=1}^n\sinh(\lambda_{n+1}-\lambda_a-{i\zeta})
\sinh(\lambda_a+\frac{i\zeta}2)}
\right)}\non
&&{\dis\hspace{27mm}
\times \hat W_n(\{\lambda\},\{z\})
\cdot {\det}_{n+1} \hat M_{jk}\cdot
{\det}_{n+1}\left[\rho(\lambda_j,z_k)\right].}
\ea
At this stage, we can again use the fact that  $\hat M$
and $(\rho(\lambda_j,z_k))$ become Cauchy matrices at $\zeta=\pi/2$. To
compute their determinants, it is convenient to use the following
modification of \eq{SZCauchy}:
\be{SSCauchy}
{\det}_n\frac1{\sinh2(x_j-y_k)}=
\frac{\prod\limits_{j>k}^n\cosh(x_j-x_k)\cosh(y_k-y_j)}
{2^{n}\prod\limits_{j,k=1}^n\cosh(x_j-y_k)}\cdot
{\det}_n\frac1{\sinh(x_j-y_k)}.
\ee
Substituting the corresponding expressions of ${\det}_{n+1}\hat M_{jk}$
and ${\det}_{n+1} \rho(\lambda_j,z_k)$ into \eq{SSs+s-new},
we arrive at
\ba{SSs+s-mod}
&&{\dis\hspace{-7mm}
\langle \sigma^+_1\sigma^-_{m+1}\rangle=\frac1{2i}\sum_{n=0}^{m-1}
\frac{1}{n!(n+1)!}
\left(\frac{i}\pi\right)^{n+1}
\int\limits_{\mathbb{R}} d^{n+1}\lambda
\prod_{a=1}^{n}\varphi^{-m+1}(\lambda_a)\cdot
\frac{1}{\sinh(\lambda_{n+1}+\frac{i\pi}{4})}}\non
&&{\dis\hspace{-7mm}
\times
\oint\limits_{C_z}
\prod_{j=1}^{n+1}\frac{dz_j}{2\pi i}
\prod_{a=1}^{n+1} \varphi^{m-1}(z_a)\cdot
{\det}_{n+1}\left(\frac1{\sinh(\lambda_j-z_k)}\right)
{\det}_{n+1}\left(
\begin{array}{c}
\frac1{\sinh(z_k-\lambda_1)}\\
\vdots\\
\frac1{\sinh(z_k-\lambda_n)}\\
\frac1{\sinh(z_k+\frac{i\pi}4)}
\end{array}\right).}
\ea

As we have seen above, the correlation function
$\langle \sigma^z_1\sigma^z_{m+1}\rangle$ and the emptiness formation
probability $\tau(m)$ at $\Delta=0$ are completely described by only one
or two terms at $\Delta=0$. The peculiarity of the correlation
function $\langle \sigma^+_1\sigma^-_{m+1}\rangle$ is that, even in
the limit of free fermions, all the terms of the corresponding series
survive. Nevertheless, it is possible to express \eq{SSs+s-mod} into a 
more compact form.

The contour integrals with respect to $z_k$ in \eq{SSs+s-mod} can be
easily computed. Due to the symmetry of the integrand with respect to
all the variables $z_k$, $1\le k \le n+1$, we can make the replacement
$$
{\det}_{n+1}\left(\frac1{\sinh(\lambda_j-z_k)}\right)\longrightarrow
(n+1)!\prod_{a=1}^{n+1}\frac1{\sinh(\lambda_a-z_a)}.
$$
Then, inserting for each $z_k$ the factors
$\sinh^{-1}(\lambda_k-z_k)$ and $\varphi^{m-1}(z_k)$ into the $k$-th
column of the remaining determinant, we can integrate separately each 
of these columns with respect to $z_k$, using the method described in
Section \ref{SZSZ}:
\be{SSs+s-modin}
\langle \sigma^+_1\sigma^-_{m+1}\rangle=\frac1{2i}\sum_{n=0}^{m-1}
\frac{1}{n!} \left(\frac{i}\pi\right)^{n+1}
\int\limits_{\mathbb{R}} d^{n+1}\lambda
\prod_{a=1}^{n}\varphi^{-m+1}(\lambda_a)
\frac{\det_{n+1}U_{jk}}{\sinh(\lambda_{n+1}+\frac{i\pi}4)},
\ee
where
\be{SSU}
U_{jk}=
\left\{ \begin{array}{ll}
{\dis \frac{1}{2\pi i}\oint\limits_{C_z}
\frac{\varphi^{m-1}(z_k) \,dz_k}
{\sinh(z_k-\lambda_j)\sinh(\lambda_k-z_k)}
=\frac{\varphi^{m-1}(\lambda_j)-\varphi^{m-1}(\lambda_k)}
{\sinh(\lambda_j-\lambda_k)},}
& {\dis\qquad j\le n,\ j\ne k,}\non
{\dis -\frac{1}{2\pi i}\oint\limits_{C_z}
\frac{\varphi^{m-1}(z_k) \,dz_k}
{\sinh^{2}(z_k-\lambda_j)}
= 2i(m-1)\frac{\varphi^{m-1}(\lambda_j)}
{\cosh(2\lambda_j)},}
& {\dis\qquad j\le n,\ j=k,}\non
{\dis
\frac{1}{2\pi i}\oint\limits_{C_z}
\frac{\varphi^{m-1}(z_k) \,dz_k}
{\sinh(z_k+\frac{i\pi}4)\sinh(\lambda_k-z_k)}
=\frac{\varphi^{m-1}(\lambda_k)}
{\sinh(\lambda_k+\frac{i\pi}{4})},}
&{\dis\qquad j=n+1.}
\end{array}\right.
\ee

The sum \eq{SSs+s-modin} is very similar to the expansion of a Fredholm
determinant of a linear integral operator.
To make it more clear, let us introduce
\be{SSnot}\label{V}
V(\lambda,\mu)=
i\frac{\Bigl(\varphi(\lambda)/\varphi(\mu)
\Bigr)^{\frac{m-1}2}-
\Bigl(\varphi(\mu)/\varphi(\lambda)
\Bigr)^{\frac{m-1}2}}{\pi\sinh(\lambda-\mu)},
\ee
where $V (\lambda, \lambda)$ is defined by continuity from \eq{V}.
Then the equation \eq{SSs+s-modin} can be written as
\be{SSs+s-almfred}
\langle\sigma^+_1\sigma^-_{m+1}\rangle=
\sum_{n=0}^{m-1} \frac{1}{n!}
\int\limits_{\mathbb{R}} d^{n+1}\lambda\cdot
{\det}_{n+1}\tilde U_{jk},
\ee
with
\be{SStiU}
\tilde U_{jk}=\left\{ \begin{array}{ll}
{\dis V(\lambda_j,\lambda_k)}&
{\dis\qquad j,k\le n,}\non
{\dis V(\lambda_j,\lambda_{n+1})\cdot
\frac{\varphi^{\frac{m-1}2}(\lambda_{n+1})}
{2i\sinh(\lambda_{n+1}+\frac{i\pi}4)}}&
{\dis\qquad k=n+1,~j\le n}\non
{\dis\frac{i\varphi^{\frac{m-1}2}(\lambda_{k})}
{\pi\sinh(\lambda_{k}+\frac{i\pi}4)}
}&
{\dis\qquad j=n+1,~k\le n}\non
{\dis
\frac{\varphi^{m-1}(\lambda_{n+1})}
{2\pi\sinh^2(\lambda_{n+1}+\frac{i\pi}4)}}&
{\dis\qquad j,k=n+1,}
\end{array}\right.
\ee
It is shown in Appendix \ref{FM} that \eq{SSs+s-almfred} results
into the derivative of a Fredholm determinant:
\be{SSFrMi}
\langle\sigma^+_1\sigma^-_{m+1}\rangle=
\frac{\partial}{\partial\alpha}
\det\left(I+V(\lambda,\mu)+\frac\alpha{2\pi}R(\lambda,\mu)\right),
\ee
where $I$ denotes the identity operator and
\be{SSR}
R(\lambda,\mu)=\frac{\Bigl(\varphi(\lambda)\varphi(\mu)\Bigr)^{
\frac{m-1}2}}{\sinh(\lambda+\frac{i\pi}4)\sinh(\mu+\frac{i\pi}4)}.
\ee
The operator $I+V+\frac\alpha{2\pi}R$ acts on the real axis,
and $\alpha$ is an auxiliary parameter. Since $R(\lambda,\mu)$ is
a one-dimensional projector, the determinant in \eq{SSFrMi} is
a linear function of $\alpha$. Hence, the derivative of the determinant
does not depend on $\alpha$.
To compare \eq{SSFrMi} with the result obtained
in \cite{ColIKT93} one can make the standard change of variables
$\cosh2\lambda =\cos^{-1} p$, $\cosh2\mu =\cos^{-1} q$. Then, the
kernel $V$ and the projector $R$ become
\be{SSVR}
V(p,q)=-\frac{\sin\frac{m-1}2(p-q)}{\pi\sin\frac{1}2(p-q)},
\qquad
R(p,q)= (-1)^{m} e^{\frac{im}2(p+q)},
\ee
where the integral operator acts on the interval $[-\pi/2,\pi/2]$. Finally,
replacing $p$ with $-p$ and $q$ with $-q$, we arrive at
\be{SSFrMires}
\langle\sigma^+_1\sigma^-_{m+1}\rangle=
(-1)^m\frac{\partial}{\partial\alpha}
\det\left(I-\frac{\sin\frac{m-1}2(p-q)}{\pi\sin\frac{1}2(p-q)}
+\frac\alpha{2\pi}e^{-\frac{im}2(p+q)}\right).
\ee
%
This formula coincides with the result of \cite{ColIKT93} up to the
factor $(-1)^m$. The existence of this factor is
due to the fact that we use a different definition for the Hamiltonian, 
as it was mentioned in
the beginning of Section \ref{XXZ}.


\section{Long-distance asymptotics of
$\langle\sigma^+_1\sigma^-_{m+1}\rangle$}\label{LA}

The leading asymptotic behavior of the correlation function
$\langle\sigma^+_1\sigma^-_{m+1}\rangle$ was computed in
\cite{Wu66,Mcc68}. Later, in \cite{ColIKT93},  a Fredholm
determinant representation of the dynamic temperature
correlation function was obtained
for an arbitrary value of the external magnetic field.
The determinant \eq{SSFrMires} appears to be a particular case
of this result. To compute its asymptotic behavior at large $m$,
one can use the methods of the matrix Riemann--Hilbert problem which were
developed in \cite{ItsIKS93} to study the dynamic temperature
correlations.  However, these methods allow to find the asymptotics
of the determinant only up to a multiplicative constant, whereas
the determinant \eq{SSFrMires} can be computed explicitly as a finite
product of $\Gamma$-functions. In this section, we present the
corresponding derivation and reproduce the results of the papers
\cite{Wu66,Mcc68}.

Observe first that the kernel $V(p,q)$ \eq{SSVR} is degenerated:
\be{LAV}
V(p,q)=-\frac{\sin\frac{m-1}2(p-q)}{\pi\sin\frac12(p-q)}=
-\frac1\pi\sum_{k=1}^{m-1}e^{i(p-q)(k-\frac m2)}.
\ee
Thus, the corresponding Fredholm determinant 
can be reduced to the determinant of
a matrix of finite size. Indeed, if a kernel $K(p,q)$ has
the form $K(p,q)=\sum_{k=1}^m f_k(p)g_k(q)$, then
\be{LAform}
\det(I+K(p,q))={\det}_m(\delta_{jk}+M_{jk}),
\ee
with
\be{LAM}
M_{jk}=\int\limits_C f_j(p)g_k(p)\,dp.
\ee
Here $C$ is the contour where the  operator $I+K$ acts.
In our case, $C=[-\pi/2,\pi/2]$, and
\be{LAf,g}
\begin{array}{l}
{\dis
f_k(p)=-\frac1\pi e^{ip(k-\frac m2)},\qquad k=1,\dots,m-1,}\non
{\dis
f_m(p)=\frac{\alpha}{2\pi} e^{-ip\frac m2},}\non
{\dis
g_k(q)= e^{-iq(k-\frac m2)},\qquad k=1,\dots,m.}
\end{array}
\ee
Thus, for $j<m$,
\be{LAMjk}
M_{jk}=-\delta_{jk}-\frac2\pi\left\{
\begin{array}{ll}
0,&\qquad\mbox{for}~j-k~\mbox{even},\\
(-1)^{(j-k-1)/2}\cdot\frac1{j-k},&\qquad\mbox{for}~j-k~\mbox{odd},
\end{array}\right.
\ee
whereas the elements of the last line of the matrix $M$ are given by
\be{LAMmk}
M_{mk}=\frac{\alpha}\pi\left\{
\begin{array}{ll}
0,&\qquad\mbox{for}~k~\mbox{even},\\
(-1)^{(k-1)/2}\cdot\frac1{k},&\qquad\mbox{for}~k~\mbox{odd}.
\end{array}\right.
\ee
Note that the parameter $\alpha$ enters only the last line. Hence,
taking the derivative of the determinant with respect to $\alpha$, we
need to differentiate only the elements of this line, and we
obtain
\be{LAnewrep}
\langle\sigma^+_1\sigma^-_{m+1}\rangle
=-\frac{2^{m-1}}{\pi^m}{\det}_m(a_{jk}),
\ee
where
\be{LAajk}
a_{jk}=
\left\{
\begin{array}{ll}
0,&\qquad\mbox{for}~j-k~\mbox{even},\\
(-1)^{(j-k-1)/2}\cdot\frac1{j-k},&\qquad\mbox{for}~j-k~\mbox{odd},
\end{array}\right.
\ee
for $j<m$, and
\be{LAamk}
a_{mk}=\left\{
\begin{array}{ll}
0,&\qquad\mbox{for}~k~\mbox{even},\\
(-1)^{(k-1)/2}\cdot\frac1{k},&\qquad\mbox{for}~k~\mbox{odd},
\end{array}\right.
\ee
for $j=m$.

Our goal is now to compute ${\det}_m(a_{jk})$. To do this, let us
first reorder the columns and the lines of the matrix $a_{jk}$ such that
it becomes a $2\times2$ block-matrix. One has to move:
\begin{enumerate}
\item
the columns with the number $2k$ to the position $k$
for $k=1,\dots,\left[\frac{m}{2}\right]$;
\item
the lines with the number $2j-1$ to the position $j$
for $j=1,\dots,\left[\frac{m}{2}\right]$;
\item
the last line with the number $m$ to the position
$\left[\frac{m}{2}\right]+1$.
\end{enumerate}
Here, $\left[\frac{m}{2}\right]$ denotes the integer part of $m$.
After these transformations, we arrive at
\be{LAblock}
{\det}_m(a_{jk})=(-1)^{m-1}\det\left(
\begin{tabular}{c|c}
${\dis a_{2j-1,2k}\quad}$ & ${\dis \;0\vphantom{\sum_k}}$\\
\hline
${\dis 0\quad}$ & ${\dis \;a_{2j-2,2k-1} \rule{0pt}{15pt}}$
\end{tabular}\right),\quad
\mbox{with}\quad a_{0,2k-1}\equiv a_{m,2k-1}.
\ee
Hereby the sizes of the blocks are $\left[\frac{m}{2}\right]
\times\left[\frac{m}{2}\right]$ for $a_{2j-1,2k}$ and
$\left[\frac{m+1}{2}\right]\times\left[\frac{m+1}{2}\right]$
for $a_{2j-2,2k-1}$. Observe now that
\be{LAb}
a_{2j-1,2k}=a_{2j-2,2k-1}=\frac{(-1)^{j-k-1}}{2j-2k-1}.
\ee
Hence, we obtain
\be{LAdeta1}
{\det}_m(a_{jk})=
-{\det}_{\left[\frac{m}{2}\right]}
\left(\frac{1}{2j-2k-1}\right)\cdot
{\det}_{\left[\frac{m+1}{2}\right]}
\left(\frac{1}{2j-2k-1}\right).
\ee
It remains to use the analog of \eq{SZCauchy} for rational
functions
\be{LACauchy}
{\det}_n\frac1{x_j-y_k}=
\frac{\prod\limits_{j>k}^n(x_j-x_k)(y_k-y_j)}
{\prod\limits_{j,k=1}^n(x_j-y_k)},
\ee
which gives
\be{LAdetb}
{\det}_m(a_{jk})=(-1)^{m-1}
\prod_{k=1}^{\left[\frac{m}{2}\right]}
\prod_{j=1\atop{j\ne k}}^{\left[\frac{m}{2}\right]}
\frac{j-k}{j-k-\frac12}
\prod_{k=1}^{\left[\frac{m+1}{2}\right]}
\prod_{j=1\atop{j\ne k}}^{\left[\frac{m+1}{2}\right]}
\frac{j-k}{j-k-\frac12}.
\ee
After the computation of the products with respect to $j$,
we substitute the result into \eq{LAnewrep} and eventually
obtain
\be{LASS}
\langle\sigma_1^+\sigma_{m+1}^-\rangle=\frac{(-1)^m}2
\prod_{k=1}^{\left[\frac{m}2\right]}
\frac{\Gamma^2(k)}{\Gamma(k-\frac12)\Gamma(k+\frac12)}
\prod_{k=1}^{\left[\frac{m+1}2\right]}
\frac{\Gamma^2(k)}{\Gamma(k-\frac12)\Gamma(k+\frac12)}.
\ee

Thus, we have computed the Fredholm determinant \eq{SSFrMires} as a
finite product of $\Gamma$-functions. This form enables one to
evaluate the large $m$ asymptotic behavior of the correlation function
$\langle\sigma_1^+\sigma_{m+1}^-\rangle$ in a rather simple way (see
Appendix \ref{AS}). The result reads
\be{LAresas}
\langle\sigma_1^+\sigma_{m+1}^-\rangle=
\frac{(-1)^m}{\sqrt{2m}}\exp\left\{
\frac12\int_0^\infty\frac{dt}t\left[e^{-4t}-
\frac1{\cosh^2t}\right]\right\}
\left(1-\frac{(-1)^m}{8m^2}+{\cal O}(m^{-4})\right).
\ee
%

\section*{Conclusion}

To conclude this article, we would like to discuss the significance of our 
results and the perspectives they open concerning the 
computation of correlation functions in a more general case. 
Actually, the purpose of the work presented here was double. 

Our first goal was to demonstrate that it was really possible 
in the free fermion limit to compute the spin-spin correlation functions 
using their multiple integral representations. Indeed, the ability of
this method to provide effective results, even in the simplest case 
$\Delta=0$, has for a long time been seriously under question.
Thanks to the new formulas obtained in
\cite{KitMST02}, which correspond in fact to certain re-summations of
the elementary blocks \cite{JimMMN92}--\cite{KitMT00}, we were able here
to solve this problem.

The second goal of this paper concerns the possible application of these new
integral representations to the general $XXZ$ model. We hope that
some of the technical methods presented here are not specific to the free
fermion point, and that they can also be successfully adapted to study a more 
general case. 

In particular, the method we used here to compute the $z$-integrals 
might be quite efficient for the evaluation of the long distance asymptotics 
of the spin-spin correlation functions. Recall that, in this paper, we
have deformed the original contour $C_z$ into a horizontal strip $\Gamma$
of width $i\pi$. At $\Delta=0$, the contribution coming from
$\Gamma$ vanishes due to the periodicity of the integrand. In the general
case this property is no longer valid since the density function
$\rho(\lambda,z)$ is no longer $i\pi$-periodical
(except at $\zeta=\frac{\pi}{2n}$, where $n$ is a positive integer).
Nevertheless, it seems that one can control the order of 
the contribution coming from $\Gamma$ as $m\to\infty$. 
Indeed, it is possible to choose the strip
such that $|\varphi(z)|<1$ (see \eq{SZphi}) on its boundaries.
Then the factor $\varphi^m(z)$ becomes exponentially small uniformly
on an arbitrary finite interval of the new integration contour
for $z$. Preliminary estimates show that the integrals over $\Gamma$
decrease as some negative powers of $m$ as $m\to\infty$. Thus, it is
very possible that the leading long distance asymptotics of
the spin-spin correlation functions are given by the residues of the
integrand within the strip $\Gamma$.

Finally, we would also like to draw the reader's attention on 
the method used for the evaluation of 
the long-distance asymptotics of the emptiness formation
probability. In fact, the saddle point approach can be applied
directly to the representation \eq{EFPtausym} in the general case as well. 
One can thus expect
that the emptiness formation probability decreases as $\exp
\{-c(\Delta,h)m^2\}$ as $m\to\infty$. The main obstacle to find the
coefficient $c(\Delta,h)$ is related to the asymptotic analysis of
${\det}_m~t(\lambda_j,\xi_k)$ (recall that for $\Delta=0$ this
determinant is explicitly computable). The problem we mention here
coincides one to one with the problem of the computation of the partition
function of the six-vertex model with domain wall boundary conditions
\cite{IzeK85}, \cite{Ize87}, which was solved for the homogeneous case 
in \cite{KorZ00}, \cite{Zin00}. For our purpose, it would be
desirable to extend these results to the inhomogeneous case as
well, which is still an open problem. It seems nevertheless that
the emptiness formation probability admits a  
Gaussian behavior.

\section*{Acknowledgments}
N. K. would like to thank the University of York, the SPhT in
Saclay and JSPS  for financial support. 
N. S. is supported by the grants INTAS-99-1782, Leading Scientific 
Schools 00-15-96046, the Programm Nonlinear Dynamics and Solitons 
and by CNRS. 
J.M. M. is supported by CNRS. 
V. T. is supported by the DOE grant DE-FG02-96ER40959 and by CNRS.  
N. K, N. S. and V. T. would like to thank the Theoretical Physics 
group of the Laboratory of Physics at ENS Lyon for hospitality, 
which made this collaboration possible.
%

\appendix
\section{Fredholm determinant}\label{FM}

Let us first recall the definition of a Fredholm determinant: if
an operator $I+K$ acts on an interval $C$ as
\be{FMact}
[(I+K)\phi](\lambda)=\phi(\lambda)+\int\limits_C
K(\lambda,\mu)\phi(\mu)\,d\mu,
\ee
then its Fredholm determinant is
\be{FMdef}
\det(I+K)=\sum_{n=0}^\infty\frac1{n!}
\int\limits_C d\lambda_1\cdots d\lambda_n\cdot
{\det}_n K(\lambda_j,\lambda_k).
\ee

The series \eq{SSs+s-almfred}  has almost the same form. To make it
exactly the same, we observe that $\rank\tilde U_{jk}\le m-1$, and hence
${\det}_{n+1}\tilde U_{jk}=0$ as soon as $n>m-1$. Therefore the sum in
\eq{SSs+s-almfred} can be extended up to infinity. 
We can then use the identity
\be{FMident}
{\det}_{n+1}W_{jk}=\left(\left.
W_{n+1,n+1}-\frac{\partial}{\partial\alpha}\right)
{\det}_n(W_{jk}+\alpha W_{j,n+1}W_{n+1,k})\right|_{\alpha=0}
\ee
which holds for any arbitrary matrix $W$.
The sum \eq{SSs+s-almfred} can thus be written as
\ba{FMfredser}
&&{\dis\hspace{-8mm}
\langle\sigma^+_1\sigma^-_{m+1}\rangle=
\sum_{n=0}^{\infty} \frac{1}{n!}
\left(\int\limits_{\mathbb{R}}
\frac{\varphi^{m-1}(\lambda_{n+1})\,d\lambda_{n+1}}
{2\pi\sinh^2(\lambda_{n+1}+\frac{i\pi}4)}
-\frac{\partial}{\partial\alpha}\right)}\non
&&{\dis\hspace{-8mm}
\times\left.\int\limits_{\mathbb{R}} d^{n}\lambda\cdot
{\det}_{n}\left(V(\lambda_j,\lambda_k)+\frac\alpha{2\pi}
\int\limits_{\mathbb{R}}V(\lambda_j,\lambda_{n+1})
\frac{\varphi^{\frac{m-1}2}(\lambda_{n+1})
\varphi^{\frac{m-1}2}(\lambda_{k})\,d\lambda_{n+1}}
{\sinh(\lambda_{n+1}+\frac{i\pi}4)\sinh(\lambda_{k}+\frac{i\pi}4)}
\right)\right|_{\alpha=0}.}
\ea
The series \eq{FMfredser} is now exactly of the form \eq{FMdef}, which
means that the correlation function
$\langle\sigma^+_1\sigma^-_{m+1}\rangle$ can be represented as the
following Fredholm determinant:
\ba{FMfred}
&&{\dis\hspace{-2mm}
\langle\sigma^+_1\sigma^-_{m+1}\rangle=
\left(\int\limits_{\mathbb{R}}
\frac{\varphi^{m-1}(\nu)\,d\nu}
{2\pi\sinh^2(\nu+\frac{i\pi}4)}
-\frac{\partial}{\partial\alpha}\right)}\non
&&{\dis\hspace{12mm}
\times\left.
\det\left(I+V(\lambda,\mu)+\frac\alpha{2\pi}
\int\limits_{\mathbb{R}}V(\lambda,\nu)
\frac{\varphi^{\frac{m-1}2}(\nu)
\varphi^{\frac{m-1}2}(\mu)\,d\nu}
{\sinh(\nu+\frac{i\pi}4)\sinh(\mu+\frac{i\pi}4)}
\right)\right|_{\alpha=0}.}
\ea

To reduce this determinant to the form \eq{SSFrMi}, we use the following
lemma:

\begin{lemma}
Let an integral operator
$$
I+K(\lambda, \mu)+\alpha
\int\limits_C K(\lambda,\nu)y(\nu)x(\mu)\,d\nu
$$
acts on an interval $C$. Hereby the kernel $K(\lambda,\mu)$ and the functions
$x(\lambda)$, $y(\lambda)$ are such that the Fredholm determinant of this
operator exists. Then
\ba{FMid}
&&{\dis\hspace{-2mm}
\left(\int\limits_{C}
x(\nu)y(\nu)\,d\nu
-\frac{\partial}{\partial\alpha}\right)\left.
\det\left(I+K(\lambda, \mu)+
\alpha\int\limits_C K(\lambda,\nu)y(\nu)x(\mu)
\,d\nu\right)\right|_{\alpha=0}
}\non
&&{\dis\hspace{22mm}
=\frac{\partial}{\partial\alpha}
\det\Bigl(I+K(\lambda, \mu)+
\alpha y(\lambda)x(\mu)\Bigr).}
\ea
\end{lemma}

{\sl Proof.}~~ Assume first that $\det(I+K)\ne0$, i.e. that there exists the
inverse operator $I-G=(I+K)^{-1}$. Then one can extract the
determinant of the operator $I+K$:
\ba{FM1trans}
&&{\dis\hspace{-2mm}
\left(\int\limits_{C}
x(\nu)y(\nu)\,d\nu
-\frac{\partial}{\partial\alpha}\right)\left.
\det\left(I+K(\lambda, \mu)+
\alpha\int\limits_C K(\lambda,\nu)y(\nu)x(\mu)
\,d\nu\right)\right|_{\alpha=0}
}\non
&&{\dis\hspace{-2mm}
=\det(I+K)\left(\int\limits_{C}
x(\nu)y(\nu)\,d\nu
-\frac{\partial}{\partial\alpha}\right)\left.
\det\left(I+
\alpha\int\limits_C G(\lambda,\nu)y(\nu)x(\mu)\,d\nu\right)
\right|_{\alpha=0}.}
\ea
The operator $\int_C G(\lambda,\nu)y(\nu)x(\mu)\,d\nu$ is a
one-dimensional projector, hence,
\be{FMdetonedim}
\det\left(I+
\alpha\int\limits_C G(\lambda,\nu)y(\nu)x(\mu)\,d\nu\right)=
1+\alpha\int\limits_C G(\mu,\nu)y(\nu)x(\mu)\,d\nu\,d\mu.
\ee
Substituting this into \eq{FM1trans}, we obtain
\ba{FM2trans}
&&{\dis\hspace{-2mm}
\left(\int\limits_{C}
x(\nu)y(\nu)\,d\nu
-\frac{\partial}{\partial\alpha}\right)\left.
\det\left(I+K(\lambda, \mu)+
\alpha\int\limits_C K(\lambda,\nu)y(\nu)x(\mu)
\,d\nu\right)\right|_{\alpha=0}
}\non
&&{\dis\hspace{-2mm}
=\det(I+K)\int\limits_{C}
\Bigl(\delta(\nu-\mu)-G(\mu,\nu)\Bigr)
y(\nu)x(\mu)\,d\nu\,d\mu.}
\ea
In its turn the last equality is equivalent to
\ba{FM3trans}
&&{\dis\hspace{-2mm}
\left(\int\limits_{C}
x(\nu)y(\nu)\,d\nu
-\frac{\partial}{\partial\alpha}\right)\left.
\det\left(I+K(\lambda, \mu)+
\alpha\int\limits_C K(\lambda,\nu)
y(\nu)x(\mu)\,d\nu\right)\right|_{\alpha=0}
}\non
&&{\dis\hspace{-2mm}
=\det(I+K)\frac{\partial}{\partial\alpha}
\det\left(I+\alpha\int\limits_{C}
(I-G)(\lambda,\nu)y(\nu)x(\mu)\,d\nu\,d\mu\right).}
\ea
Finally, transforming the product of the two determinants into
a single one, we arrive at \eq{FMid}.

If $\det(I+K)=0$, we can consider the modified operator $I+\gamma K$,
where $\gamma$ is some complex. The Fredholm determinant
$\det(I+\gamma K)$ is an entire function of $\gamma$ and, hence,
it has a finite number of isolated zeros in any closed domain of the
complex plane. Thus, we can choose $\gamma$ such that
$\det(I+\gamma K)\ne 0$, repeat all the transformations described
above and continue the result to the point $\gamma=1$.
Thus, the lemma is proved.
\qed

It remains to observe that the structure of the determinant
\eq{FMfred} coincides with the one of \eq{FMid} if we set
$K(\lambda,\mu)=V(\lambda,\mu)$ and
\be{FMxy}
x(\lambda)=y(\lambda)=
\frac{\varphi^{\frac{m-1}2}(\lambda)}
{\sqrt{2\pi} \sinh(\lambda+\frac{i\pi}4)}.
\ee
Thus, we arrive at \eq{SSFrMi}.


\section{Asymptotic study of the product of $\Gamma$-functions}\label{AS}

We need to compute the asymptotic behavior of the quantity
\be{ASphi}
e^{-\phi_N}=\prod_{k=1}^N
\frac{\Gamma^2(k)}{\Gamma(k-\frac12)\Gamma(k+\frac12)}
\ee
where $N\to\infty$. Expanding
$\phi_N$ into a Tailor series, we obtain
\be{ASexpan}
\phi_N=2\sum_{k=1}^N\sum_{n=1}^\infty\frac1{(2n)!}\left(\frac12\right)^{2n}
\psi^{(2n-1)}(k),
\ee
where
\be{ASpsi}
\psi^{(2n-1)}(z)=\frac{d^{2n}}{dz^{2n}}\log\Gamma(z).
\ee
The sum with respect to $k$ in \eq{ASexpan} can be computed using
\be{another}
\sum_{k=1}^N\psi^{(s)}(k)= N\psi^{(s)}(N+1)+s\left(\psi^{(s-1)}(N+1)-
\psi^{(s-1)}(1)\right).
\ee
Substituting this into \eq{ASexpan}, we obtain
\be{ASexpannew}
\phi_N=2\sum_{n=1}^\infty\frac1{(2n)!}\left(\frac12\right)^{2n}
\left[N\psi^{(2n-1)}(N+1)+(2n-1)\left(\psi^{(2n-2)}(N+1)-
\psi^{(2n-2)}(1)\right)\right].
\ee
This series is absolutely convergent, and we can make an asymptotic estimate
of each term separately. To do this, let us present $\phi_N$ in the form
\be{ASphiform}
\phi_N=\frac14\log N + S_1+S_2,
\ee
where
\be{ASS1}
S_1=\frac14(N\psi'(N+1)+\psi(N+1)+{\bf C}-\log N)-
2\sum_{n=2}^\infty\frac{2n-1}{(2n)!}\left(\frac12\right)^{2n}
\psi^{(2n-2)}(1).
\ee
Here ${\bf C}=-\psi(1)$ is the Euler constant. The last term
in \eq{ASphiform} is
\be{ASS2}
S_2=2\sum_{n=2}^\infty\frac1{(2n)!}\left(\frac12\right)^{2n}
\left[N\psi^{(2n-1)}(N+1)+(2n-1)\psi^{(2n-2)}(N+1)\right].
\ee
Recall also the asymptotic expansion for the logarithm of
$\Gamma$-function
\be{ASexplog}
\log\Gamma(z)=z\log z-z-\frac12\log z+\frac12\log(2\pi)
+\sum_{k=1}^{n-1} \frac{B_{2k}}{2k(2k-1)z^{2k-1}}+
{\cal O}(z^{1-2n}),\qquad |z|\to\infty,
\ee
where $B_{2k}$ are Bernoulli numbers. Using \eq{ASexplog} one can easily
see that at $N\to\infty$ the term $S_1$ has a finite limit, while
$S_2$ vanishes. 

To compute the limiting value of $S_1$, one can use
for example the integral representation for the derivatives of
$\psi(z)$:
\be{ASintrep}
\psi^{2n-1}(z)=\int_0^\infty
\frac{e^{-tz}t^{2n-1}}{1-e^{-t}}\,dt,\qquad
n\ge1,\qquad \Re(z)>0.
\ee
Then the series in \eq{ASS1} becomes the expansion of the exponent, and
we obtain
\be{ASest}
S_1=-\frac14
\int_0^\infty\frac{dt}t\left[e^{-4t}-
\frac1{\cosh^2t}\right],
\qquad N\to\infty.
\ee
This gives us constant the contribution to $\phi_N$.

In order to find the corrections to this quantity, we need first to
take into account the higher order corrections to $\psi'(N+1)$
and $\psi(N+1)$ in $S_1$ and then to take several terms from the
series $S_2$. In particular, we find
\be{ASasphi}
\phi_N=\frac14\log N-\frac14
\int_0^\infty\frac{dt}t\left[e^{-4t}-
\frac1{\cosh^2t}\right]+\frac{1}{64N^2}
+{\cal O}(N^{-4}),
\qquad N\to\infty.
\ee
This formula can be directly applied for the computation
of the asymptotic behavior of \eq{LASS}.

\end{document}